\documentclass[pre,floatfix,twocolumn,showpacs]{revtex4}
\usepackage{amsmath}
\usepackage{amssymb}
\usepackage{graphicx}

\newcommand{\eg}{\emph{e.}$\,$\emph{g.}}
\newcommand{\ie}{\emph{i.}$\,$\emph{e.}}
\newcommand{\etal}{\emph{et}$\,$\emph{al.}}

\newcommand{\romb}{{\operatorname{b}}}
\newcommand{\romc}{{\operatorname{c}}}
\newcommand{\romd}{{\operatorname{d}}}
\newcommand{\rome}{{\operatorname{e}}}
\newcommand{\romg}{{\operatorname{g}}}
\newcommand{\romi}{{\operatorname{i}}}
\newcommand{\romm}{{\operatorname{m}}}
\newcommand{\romB}{{\operatorname{B}}}
\newcommand{\romlg}{{\operatorname{lg}}}

\newcommand{\romf}{{\operatorname{f}}}

\newcommand{\romtot}{{\operatorname{tot}}}

\newcommand{\VECa}{{\boldsymbol{a}}}

\newcommand{\VECn}{{\boldsymbol{n}}}
\newcommand{\VECq}{{\boldsymbol{q}}}
\newcommand{\VECr}{{\boldsymbol{r}}}
\newcommand{\VECu}{{\boldsymbol{u}}}

\newcommand{\sinc}{{\operatorname{sinc}}}

\newcommand{\tR}{\tilde{R}}

\newcommand{\ZZ}{\mathbb{Z}}

\begin{document}

\title{Solvent free model for self-assembling fluid bilayer membranes:\\ Stabilization of the fluid phase based on broad attractive tail potentials}

\author{Ira R. Cooke}
\author{Markus Deserno}

\affiliation{Max-Planck-Institut f\"ur Polymerforschung, %
             Ackermannweg 10, %
             55128 Mainz, %
             Germany}

\date{\today}
\begin{abstract}
  We present a simple and highly adaptable method for simulating
  coarse-grained lipid membranes without explicit solvent.  Lipids are
  represented by one head-bead and two tail-beads, with the
  interaction between tails being of key importance in stabilizing the
  fluid phase.  Two such tail-tail potentials were tested, with the
  important feature in both cases being a variable range of
  attraction.  We examined phase diagrams of this range versus
  temperature for both functional forms of the tail-tail attraction
  and found that a certain threshold attractive width was required to
  stabilize the fluid phase.  Within the fluid phase region we find
  that material properties such as area per lipid, orientational
  order, diffusion constant, inter-leaflet flip-flop rate and bilayer
  stiffness all depend strongly and monotonically on the attractive
  width.  For three particular values of the potential width we
  investigate the transition between gel and fluid phases via heating
  or cooling and find that this transition is discontinuous with
  considerable hysteresis.  We also investigated the stretching of a
  bilayer to eventually form a pore and found excellent agreement with
  a recently published analytic theory.
\end{abstract}

\pacs{61.20.Ja, 81.16.Dn, 82.70.Uv}

%
%
%

\maketitle


\section{Introduction}

Lipid bilayers are among the most versatile of nature's biomaterials.
As the interface between the cell and its environment, or between
organelles and the cytosol, they provide regulated transport of
substances as small as protons to as large as entire cells.  Such a
wide range of functional length scales is naturally studied via an
equally broad range of methods.  For example, at the smallest scale
detailed quantum atomistic simulations are necessary to study the
transport of ions or water across the membrane interface
\cite{zahn_brickmann:2001,smondyrev_voth:2002}, whereas at the
opposite end of the length scale spectrum, analytic theory
\cite{seifert:1997} and dynamically triangulated lattice simulations
\cite{kumar_etal:2001} have been used to determine the shape behavior
of whole vesicles under various conditions of pressure, volume and
area, or even under hydrodynamic flow
\cite{gompper_noguchi:2004}. Between these two extremes are problems
at the so called ``meso-scale'' which include membrane fusion and
rupture
\cite{shillcock_lipowsky:2005,noguchi_takasu:2001a,mueller_etal:2003},
domain formation in multi-component membranes
\cite{veatch_keller:2002,veatch_keller:2003,baumgart_etal:2003,laradji_kumar:2004},
the coupling of membrane composition with curvature, or the
interaction of membranes with colloidal or viral particles
\cite{noguchi_takasu:2002a}.  Such problems occur at relatively large
length and timescales but at the same time require a particle based
approach that reproduces the basic bilayer structure of the membrane.
The combination of these two requirements necessitates the use of
coarse grained simulation approaches, in which groups of atoms are
represented by single particles.  Such coarse grained approaches vary
in their level of detail from a single bead per lipid
\cite{drouffe_etal:1991} up to quite detailed lipids with on the order
of ten beads \cite{marrink_etal:2004}.  Naturally such a range of
levels of coarse graining goes along with a tradeoff between
computational efficiency and level of detail.  In this respect, the
single most important determinant of model speed appears to be the
presence or absence of explicit solvent.  Unlike the two-dimensional
membrane, the solvent is the \emph{bulk} phase which fills the entire
simulation box, and integrating its degrees of freedom can easily
amount to more than 90\% of simulation time.  Naturally, one must
include solvent when its effects are of inherent interest to the
physics of the problem.  In many cases however, its task is merely to
mediate the hydrophobic attraction between lipid tails, and as such it
is secondary to the overall purpose of the simulation.

If one could simulate bilayer membranes without the need for explicit
solvent, a vast increase in accessible length and time scales would
result, yet despite the long and successful history of solvent free
models in polymer physics, this approach has not yet been widely
adopted for lipid bilayer simulations.  This is because the membrane
case displays one additional complication: Unlike polymers, whose
structure is at the outset determined by \emph{chemistry}, lipids
first have to \emph{physically} self-assemble into a two-dimensional
fluid bilayer.  This aggregation results from a balance between lipid
\emph{entropy} and the \emph{energy} of cohesion.  Since in the
solvent-free case the latter stems from effective attractions (for
instance between lipid tails), a physically meaningful balance will
pose restrictions on the interaction potentials.  Indeed, the
collective experience from the past has shown that simple choices
(\eg\ Lennard-Jones, LJ) do not lead to a fluid bilayer phase but only
to ``solid'' bilayers at low temperature and low density (``gas'')
phases at high temperature.  Concluding that simple pair potentials
are insufficient, researchers have then turned to the use of density
dependent (multibody) interactions
\cite{drouffe_etal:1991,noguchi_takasu:2001a,noguchi_takasu:2001b,noguchi_takasu:2002a,noguchi_takasu:2002b},
angular dependent potentials \cite{brannigan_brown:2004} or highly
tuned sets of Lennard-Jones like potentials \cite{farago:2003} to
stabilize the fluid bilayer phase without solvent (see Brannigan
\etal\ for a recent review \cite{brannigan_etal:2005r}).
Unfortunately, each of these approaches suffers from one or more
significant drawbacks.  For example, the multibody approach introduces
complications for interpretation and measurement of thermodynamic
quantities, while neither the angular dependent approach nor the use
of tuned LJ potentials has so far led to bilayers for which unassisted
self-assembly has been demonstrated.  In addition to these technical
problems, it is also notable that the reported bending stiffnesses are
generally outside the experimentally reported range
\cite{evans_rawicz:1990,seifert_lipowsky:1995,rawicz_etal:2000}, being
restricted to either relatively low  (\cite{drouffe_etal:1991,noguchi_takasu:2002b} $<5 \,k_\romB T$) or  high (\cite{farago:2003,brannigan_brown:2004} $> 50\,k_\romB T$) values.
Thus, there remains a clear need for an efficient solvent-free bilayer
model that does not suffer from such technical drawbacks and is also
highly tuneable.

Recently, two new solvent free models have appeared which set out to
solve these problems.  One has been proposed by Brannigan
\etal\ \cite{brannigan_etal:2005n}, the other one by us
\cite{cooke_etal:2005}.  Both models display a wide parameter range in
which simple pair attractions drive an unassisted self-assembly into a
fluid phase, within which the bending stiffness can be easily tuned.
They differ from all previous solvent free models in their use of
attractive potentials that extend somewhat further than a simple
LJ-potential.  These act either between all tail beads
\cite{cooke_etal:2005} or are restricted to special interface beads
between hydrophilic head and hydrophobic tail
\cite{brannigan_etal:2005n}.  As we shall show in Sec.~\ref{sec:pd},
our experience with broad attractions of various functional forms
strongly suggests that it is this feature which ultimately enables a
fluid bilayer phase for these strongly coarse grained systems.

Although similar in spirit, these two models differ in a number of
details.  For instance, Brannigan \etal\ opt for a more detailed
representation of lipid molecules compared to us, with the intention
to capture local features of the bilayer stress profile more
accurately, but at a concomitant price in efficiency.  It therefore
depends on the physical problem under study, as well as on ones
position in the detail \emph{vs} efficiency tradeoff, which model is
preferable in any given situation.  A judicious choice then requires
good knowledge of the physical properties of these models, which have
so far been outlined only rather briefly, and the present paper goes
toward filling this gap for the model developed by us.  In
Sec.~\ref{sec:basic} and \ref{sec:pd} we briefly describe our model
and discuss its self-assembly properties.  In particular, we support
our claim that the long-ranged nature of the attractions are the
feature of key importance by showing that the qualitative physical
properties are robust against change of the specific functional form
of the attraction.  In Sec.~\ref{sec:fluid} a detailed account of the
properties of the fluid phase and their variation with attractive
width and temperature is given along with a description of the
gel-fluid transition.  Finally, in Sec.~\ref{sec:stretch} we study the
stretching and rupture of a bilayer sheet and find near perfect
agreement with a simple theoretical model developed by Farago
\cite{farago:2003} and also Tolpekina \etal\ \cite{tolpekina_etal:2004} as
well as with experimental data.


\section{Basic Principles of the Model}\label{sec:basic}

We describe a model surfactant system which is based on the simple
idea that the solvent mediated interaction between lipid tails can be
represented by an effective attractive potential of sufficiently broad
range.  Such a potential can meet the two demands of (\emph{i})
providing enough cohesive energy to drive assembly of a
two-dimensional aggregate, while (\emph{ii}) permitting enough lateral
freedom for the lipid constituents to remain in a fluid state.  There
are many ways in which one might construct a coarse grained model
based around the central principle of such a broad attractive
potential. Below, we shall present two alternatives which differ in the
exact form of their tail attractions.  In showing both of these we
merely seek to demonstrate that it is not the precise functional form
that matters but merely the attractive range.

In this paper we present the most coarse grained version of our model,
because this is the most useful in terms of length scales obtainable.
Nonetheless, it is worth noting that the same principles can trivially
be extended to include more detailed lipids with greater numbers of
atoms should this be required.  Of course, one should always remember
that by increasing the number of lipid beads, the ratio of simulation
lengths to real lengths becomes less favorable. The lipids we
use are represented by one ``head'' bead followed by two ``tail''
beads.  Their size is fixed via a Weeks-Chandler-Andersen potential
\begin{equation}
  V_{\text{rep}}(r;b) = \left\{
  \begin{array}{c@{\;\;,\;\;}c}
    4\epsilon\big[(\frac{b}{r})^{12}-(\frac{b}{r})^6+\frac{1}{4}\big] & r \le r_\romc \\
    0 & r>r_\romc
  \end{array}
  \right. \ ,
\end{equation}
with $r_\romc=2^{1/6}b$.  We use $\epsilon$ as our unit of energy.  In
order to ensure an effective cylindrical lipid shape we choose
$b_{\text{head,head}} = b_{\text{head,tail}}=0.95\,\sigma$ and
$b_{\text{tail,tail}}=\sigma$, where $\sigma$ is our unit of length.
The three beads are linked by two FENE bonds
\begin{equation}
  V_{\text{bond}}(r) = -\textstyle\frac{1}{2}k_{\text{bond}}\,r_\infty^2\log\big[1-(r/r_\infty)^2\big] \ ,
\end{equation}
with stiffness $k_{\text{bond}}=30\,\epsilon/\sigma^2$ and divergence
length $r_\infty=1.5\,\sigma$.  Lipids are straightened by a harmonic
spring with rest length $4\sigma$ between head-bead and second
tail-bead
\begin{equation}
  V_{\text{bend}}(r) = \textstyle\frac{1}{2}k_{\text{bend}}(r-4\sigma)^2 \ ,
\end{equation}
which corresponds in lowest order to a harmonic bending potential
$\frac{1}{2} k_{\text{bend}}\sigma^2\,\vartheta^2$ for the angle
$\pi-\vartheta$ between the three beads.  We fixed the bending
stiffness at $k_{\text{bend}}=10\,\epsilon/\sigma^2$.  

The absence of explicit solvent molecules and the hydrophobic effect
they would give rise to is compensated by an attractive interaction
between all \emph{tail} beads.  We compared two alternative potentials
that account for this effect (see insets to Figs. \ref{fig:cospd} and
\ref{fig:ljpd} ).  The first of these
\begin{equation}
  \label{eqn:cos}
  V_{\text{cos}}(r) = \left\{
  \begin{array}{c@{\;\;,\;\;}c}
    -\epsilon & r < r_\romc \\
    -\epsilon\,\cos^2\frac{\pi(r-r_\romc)}{2w_\romc} & r_\romc \le r \le r_\romc + w_\romc \\
    0 & r > r_\romc+w_\romc
  \end{array}
  \right. 
\end{equation}
describes an attractive potential with depth $\epsilon$ which for
$r>r_\romc$ smoothly tapers to zero.  In this case, tuning the decay
range $w_\romc$ proves to be the key to obtaining a fluid bilayer
state.  

The second alternative is based on the familiar Lennard-Jones
potential but extends its range simply by inserting a flat piece of
length $w_\romf$ at the minimum
\begin{eqnarray}
  \label{eqn:lj}
  V_{\text{flat LJ}}(r) &=& 
\\
 & & \hspace {-5em} \left\{
  \begin{array}{c@{\;\;,\;\;}c}
    -\epsilon & r < r_\romc + w_\romf \\
    4\epsilon\big[(\frac{b}{r-w_\romf})^{12}- (\frac{b}{r-w_\romf})^6 \big] & r_\romc \le r \le w_\romf + w_{\text{cut}} \\
    0 & r > w_\romf + w_{\text{cut}}
  \end{array}
  \right. 
\nonumber
\end{eqnarray}
where our key tuning parameter is now the width $w_\romf$ of the flat
region.  The potential is cut-off beyond $w_\romf+w_{\text{cut}}$,
where $w_{\text{cut}}=2.5\,\sigma$ is set to the usual value.  Note
that tuning $w_\romf$ achieves a broad potential $V_{\text{flat LJ}}$
simply by introducing a flat region, whereas in the case of
$V_{\text{cos}}$ the tuning parameter $w_\romc$ varies decay range and
shape simultaneously.

The above model is sufficiently simple to allow implementation with a
variety of molecular dynamics (MD) integration schemes or even
Monte-Carlo.  In this work we performed MD simulations with a Langevin
thermostat to obtain the canonical ensemble \cite{GrKr86} (time step
$\delta t=0.01\,\tau$ and a friction constant $\Gamma=\tau^{-1}$ in
Lennard-Jones units \footnote{If beads have a mass $m$, the LJ time
scale is given by $\tau=\sigma\sqrt{m/\epsilon}$.}). Constant volume
simulations were performed using a cuboid box with sides $L_x = L_y$,
$L_z$ subject to periodic boundary conditions.  If needed, constant
tension conditions were also implemented via a modified Andersen
barostat \cite{kolb_duenweg:1999} allowing box resizing in $x$ and $y$
dimensions only (with a box friction $\Gamma_{\text{box}} = 2 \times
10^{-4}\,\tau^{-1}$ and box mass within the range $Q = 10^{-5} \ldots 10^{-4}$).  By employing such a barostat rather than a
simple Monte Carlo box length move we aimed to preserve the correct
fluctuation behavior of our system.

All simulations were performed using the ESPResSo program
\cite{espresso}.  Lipid membrane specific analysis and setup was done
using the mbtools package which is included as part of the main ESPResSo
distribution.


\section{Bilayer Stability and Self Assembly}\label{sec:pd}

\begin{figure}
\includegraphics[scale=0.34]{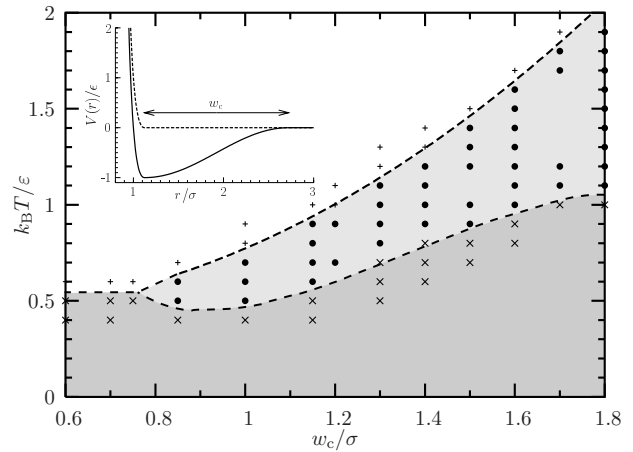}
\caption{Phase diagram resulting from $V_{\text{cos}}$ cohesion
  (Eqn.~\ref{eqn:cos}) in the plane of potential width $w_\romc$ and
  temperature at zero lateral tension.  Each symbol corresponds to one
  simulation and identifies different bilayer phases: $\times$: gel;
  $\bullet$: fluid, {\footnotesize $+$}: unstable.  Lines are merely
  guides to the eye.  The inset shows the pair-potential between tail
  lipids (solid line) and the purely repulsive head-head and head-tail
  interaction (dashed line).}\label{fig:cospd}
\end{figure}

\begin{figure}
\includegraphics[scale=0.8]{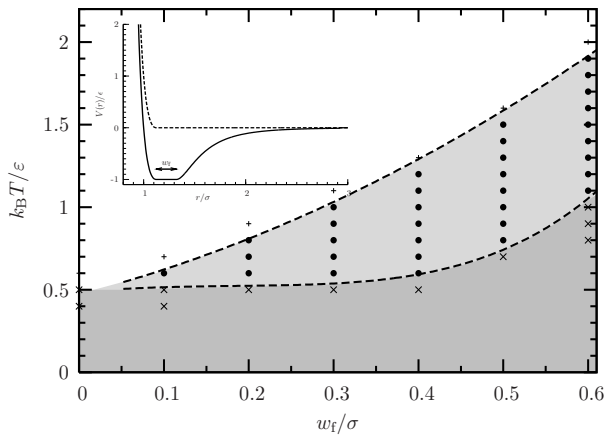}
\caption{Phase diagram resulting from $V_{\text{flat LJ}}$ cohesion
  (Eqn.~\ref{eqn:lj}) in the plane of potential width $w_\romf$ and
  temperature at zero lateral tension.  The meaning of all symbols is
  the same as for figure (Fig.~\ref{fig:cospd})}\label{fig:ljpd}
\end{figure}

In this section we shall map out the conditions under which a
tensionless fluid bilayer state is stable.  To do this we identified
the fluid phase in two different ways (see Figs.~\ref{fig:cospd} and
~\ref{fig:ljpd} for the following) .  In the first method, a
box-spanning bilayer was pre-assembled from 1000 lipids, and its
equilibration under zero lateral tension was attempted.  This resulted
in one of three possible outcomes: a stable fluid bilayer, a gel phase
with strongly increased lipid order and much lower diffusion constant,
or complete breakup of the system to form a ``gas'' phase.  The
transition between gel and liquid occurred over a narrow temperature
range and will be explored in more detail in
Sec.~\ref{subsec:gelfluid}.  The gas phase was always identified as
the point at which the imposition of zero tension conditions resulted
in a divergence in the box length.

\begin{figure}
\includegraphics[scale=1.0]{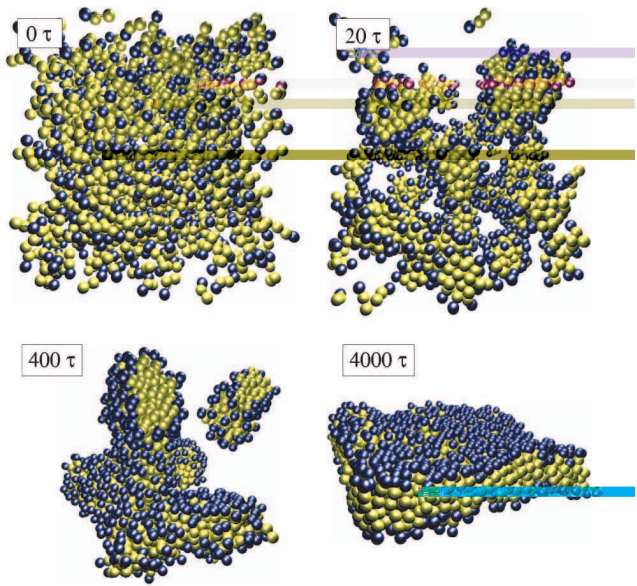}
\caption{Self-assembly sequence for the bilayer system with $1000$
  lipids in a cubic box of side length $25 \sigma$.  Lipid cohesion
  was set to $w_\romc/\sigma = 1.4$ ($V_{\text{flat LJ}}$) and
  temperature to $k_\romB T = \epsilon$.  A random gas of lipids
  quickly forms small clusters which slowly coarsen and eventually
  ``zip up'' to form a box-spanning bilayer sheet.  The numbers
  indicate the MD time.}\label{fig:zip}
\end{figure}

Our second method for identifying the fluid bilayer phase was designed
to ensure that we were not artificially stabilizing the bilayer due to
pre-assembly.  In this case we conducted constant volume simulations
starting from a random ``gas'' configuration.  Under all conditions
which previously gave stable tensionless membranes, a bilayer patch
quickly self-assembled, which, at the correct box size could zip up
with its open ends to span the box (see Fig.~\ref{fig:zip} for such a
sequence).  If the box was too big, the patch either remained free, or
(sometimes) closed upon itself to form a vesicle.  For rather large
values of the width parameter $w_\romc$ or $w_\romf$ the line between
fluid bilayer and the isotropic ``gas'' state becomes less distinct.
For example, we observed that for $w_\romc\gtrsim1.6\,\sigma$ and
$w_\romf\gtrsim 0.4\,\sigma$, self assembly to box spanning bilayers
(at constant volume) could occur well above the evaporation boundary
and that immediately below this boundary we observed rather indistinct
bilayers, with particularly high flip-flop rates and low orientational
order.  Although we will see in Sec.~\ref{sec:cross_bilayer} that these
relatively disordered states still show a definite bilayer structure,
we should not be surprised if strange behaviors are found in this
small region, and we would therefore caution against the physical
interpretation of such results without additional checks for model
artifacts.

Using the methods mentioned above, we determined the phase diagrams
for both types of attractive tail potentials, Eqns.~(\ref{eqn:cos})
and (\ref{eqn:lj}), shown in Figs.~\ref{fig:cospd} and \ref{fig:ljpd},
respectively.  The most important point to note is that in both cases
we see a fluid bilayer region that broadens significantly as potential
width is increased.  In this sense, both phase diagrams are remarkably
similar given the strong differences in the nature of the functional
forms and tuning parameters used to obtain them.  At a more detailed
level, one can see differences in the shape of the transition lines
between the two models.  This is likely due to the fact that the two
width parameters work entirely differently.  In the case of the cosine
attraction, Eqn.~(\ref{eqn:cos}), the attractive gradient is actually
varied along with $w_\romc$, whereas for Eqn.~(\ref{eqn:lj}), varying
$w_\romf$ leaves the attractive shape unchanged and merely adds a
region of zero force close to the particle.

Based on the relative similarity of our results for two such radically
different potentials, our expectation is therefore  that other
potentials with a broad attractive width should also share the
important property of exhibiting a stable fluid bilayer phase.  This
finds further support in the observation that the recent model of
Brannigan \etal\ \cite{brannigan_etal:2005n} also uses a broad $1/r^2$
attraction.  Of course, a direct comparison is difficult because they
restrict this attraction to the interface bead between lipid head and
tail.  Notwithstanding the physical motivation of this choice
as being in accord with knowledge of the lateral stress profile, it
would be interesting to check whether \emph{fluidity} is ultimately
insensitive to this detail and rests on the long range alone.

Faced with these possibilities, as well as others we might think of
for the tail attraction, the question is now which to choose.
Since our potential attempts to capture the effects of solvent
exclusion, lipid-lipid interactions, and the fact that our 3 bead
``lipid'' is a highly coarse grained representation of the real thing,
it is difficult to guess what its functional form should look like.
Instead we could attempt to differentiate between models on the basis
of their emergent physical properties; however, it turns out that both
are highly tuneable over a similar range. Therefore our primary
considerations are practical ones.  In this regard the cosine
attraction, Eqn.~(\ref{eqn:cos}), is preferable since it acts over a
slightly shorter range than the broadened LJ potential and is
therefore faster to compute. The remainder of this paper will
therefore focus on it alone.


\section{Properties of the fluid phase}\label{sec:fluid}

We have characterized bilayers in both fluid and gel phases by several
observables, including the lipid orientational order parameter,
cross-bilayer density profiles and bending modulus, as well as the
dynamical quantities diffusion constant and flip-flop-rate.  In
Sec.~\ref{sec:observables} we shall define these quantities and explain
in detail how they are measured.  In Sec.~\ref{sec:const_T} the
results are presented as cross sections at constant temperature,
while Sec.~\ref{subsec:gelfluid} examines the fluid-gel transition via cross sections at constant potential width.


\subsection{Observables}\label{sec:observables}

\subsubsection{Orientational order parameter}

Lipids in the fluid $\text{L}_\alpha$ phase are on average oriented
parallel to the bilayer normal.  The amount of alignment can be
quantified by an orientational order parameter $S$, defined by
\begin{equation}
S = \frac{1}{2} \langle 3 (\VECa_i \cdot \VECn)^2 -1 \rangle_i
\label{eq:s}
\end{equation}
where $\VECa_i$ is the unit vector along the axis of the
$i^{\text{th}}$ lipid, $\VECn$ is the average bilayer normal and
angular brackets indicate an average over all lipids. A completely
isotropic system will have $S = 0$ whereas a fully ordered crystalline
bilayer will have $S = 1$.

\subsubsection{Cross-bilayer density profile}\label{sec:cross_bilayer}

In a well-defined bilayer the lipid distribution perpendicular to the
bilayer plane is very regular.  In particular, each of the constituent
beads should occupy a well defined vertical distance from the bilayer
midplane. To investigate this aspect of fluid bilayer
structure we have calculated the number density of beads $\rho(z)$ as
a function of vertical distance $z$ from the local bilayer midplane.
We used systems with 4000 lipids at constant zero tension  with a lateral box size of $L_x=L_y\approx 50\,\sigma$.
Although such a large system provides good statistics, it also
introduces the problem of dealing with undulations.  We solved this by
first assigning lipids to a $16 \times 16$ grid in the $xy$-plane and
measuring the height $z$ with respect to the average local height for
that grid cell.  Failure to do so will overestimate the width of the
distributions significantly.

\begin{figure}
\includegraphics[scale=0.75]{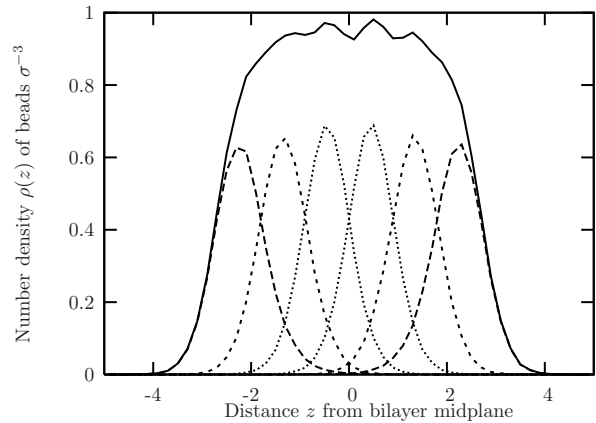}
\caption{Profile of the density $\rho$ as a function of vertical
distance $z$ from the bilayer midplane for a system of 4000 lipids at
constant zero tension and with simulation parameters $k_\romB T = 1.1
\epsilon$ and $w_\romc = 1.6$.  Plotted lines are bead densities for
head beads (dotted line), first tail beads (dot-dashed line), terminal
tail beads (dashed line), and the sum of all beads (solid line). }
\label{fig:zprof}
\end{figure}

Calculations of $\rho(z)$ for weakly coarse grained
\cite{shelley_etal:2001} or fully atomistic lipid bilayers
\cite{husslein_etal:1998} typically show differences between the shape
and width of $\rho(z)$ for each functional group in the lipid.  Of
course, a strongly coarse grained model such as ours cannot hope to
reproduce such subtleties.  Our main concern is rather to ensure that
the bilayer is not merely a loose agglomeration of lipids but that
these lipids are oriented approximately vertically and that they do
not strongly interdigitate (although such interdigitated phases do
occur naturally under certain circumstances \cite{sjripple:2005,kranenburg_smit:2005} they are not
the ``norm'' for a fluid bilayer).  From Fig.~\ref{fig:zprof} one can
verify that a well defined bilayer structure is indeed present for our
system.  Each of the three lipid beads shows a sharp peak about its
average $z$.  While the width of these peaks broadens upon approaching
the liquid-gas boundary, their location is relatively stable.  Upper
and lower head beads are separated by a distance of approximately
$4.5\,\sigma$, while the inflection points of the summed density are
separated by about $5\,\sigma$.  This agrees with our expectations for
vertically oriented lipids having only minor interdigitation.  Further evidence for
distinct bilayer leaflets can be seen in the summed bead density which
shows a slight peak for each of the terminal tail beads and a minimum
at the bilayer midplane.

In order to quantitatively examine trends in peak broadening and
overlap throughout the fluid phase region of our phase diagram we
define an overlap parameter $\Psi$ as follows:
\begin{equation}
\Psi = \frac{1}{\Omega} \sum_{i<j} \int \romd z \; \rho_i(z) \rho_j(z)
\label{eq:psi}
\end{equation}
where $i$ and $j$ are labels for the bead type (\ie, head, tail$_1$,
tail$_2$) and $\Omega = \sum_i \int \romd z \, (\rho_i(z))^2$ is a
normalization factor.  From this definition we see that complete
overlap (or bead equivalence) is indicated by $\Psi = 1$, whereas a
rigid crystalline structure with no overlap would give $\Psi = 0$.

\subsubsection{Bending modulus}\label{sec:bending_modulus}

One of the key material properties of a macroscopic membrane is its
bending stiffness, which measures the energetic cost per unit area of
imposing a local curvature.  More precisely, the classical continuum
description \cite{canham:1970,helfrich:1973} in the limit of almost
flat membranes states that the energy of a deformed piece of membrane
is given by
\begin{equation}
  E = \frac{1}{2} \int \romd x \, \romd y \; \big[ \kappa(\Delta h)^2 + \Sigma(\nabla h)^2 \big] \ ,
\label{eq:helfrich}
\end{equation}
where $\kappa$ and $\Sigma$ are bending modulus and lateral tension,
respectively, and where $h(x,y)$ describes the height of the membrane
above some reference plane (``Monge gauge'').  If we expand $h(x,y)$
in Fourier modes according to
\begin{equation}
h(\VECr)=\sum_\VECq h_\VECq \, \rome^{\romi\VECq\cdot\VECr}
\quad\text{with}\quad
\VECq=\frac{2\pi}{L}(n_x,n_y)
\end{equation}
and insert the result into Eqn.~(\ref{eq:helfrich}), we see that the
energy reduces to a sum of uncoupled harmonic oscillators with the
modes $h_\VECq$ as the degrees of freedom.  From the equipartition
theorem we then get the power spectrum of modes as \cite{seifert:1997}
\begin{equation}
  \langle |h_\VECq^2|\rangle = \frac{k_\romB T}{L^2[\kappa q^4+\Sigma q^2]} \ .
  \label{eq:hq2}
\end{equation}
It is standard practice to obtain the bending modulus from a fit of
the measured fluctuation spectrum to Eqn.~(\ref{eq:hq2}).  However,
some care has to be taken here.  First, Eqn.~(\ref{eq:helfrich}) is a
continuum description and thus requires us to focus on large length
scales.  But for wave-vectors $q$ smaller than $q_\romc =
\sqrt{\Sigma/\kappa}$ the dominant influence is the tension $\Sigma$,
and $\langle |h_\VECq^2|\rangle\sim q^{-2}$ is insensitive to
$\kappa$, so larger wave-vectors than $q_\romc$ are needed.  However,
once $1/q$ becomes comparable to the bilayer thickness, simple
continuum theory breaks down and further effects (\eg\ protrusion
modes \cite{lipowsky_grotehans:1993}) set in.  Identifying the
characteristic $q^{-4}$ scaling of the bending regime over a
sufficiently wide range thus requires $q_\romc$ -- and therefore the
lateral tension $\Sigma$ -- to be as small as possible.
Unfortunately, it turns out to be extremely hard to eliminate any
remaining tension by adjusting the simulation box size by hand, since
the tension depends very sensitively on bilayer area (see
Sec.~\ref{sec:stretch} below).  We avoided this difficulty by instead
using a modified Andersen barostat \cite{kolb_duenweg:1999} to
simulate in an ensemble of constant zero \emph{tension}.  Furthermore, to get away from microscopic
lengths we took systems four times as big as the ones we used for mapping
the phase diagram (4000 lipids, $L\simeq 50\,\sigma$).  Note that
reaching the continuum limit in MD simulations is not trivial, since
the relaxation time of bending modes scales as $q^{-4}$.

\begin{figure}
\includegraphics[scale=0.7]{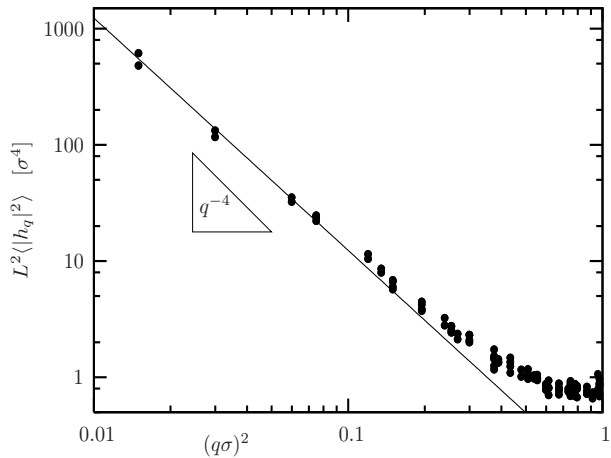}
\caption{Asymptotic $q^{-4}$ scaling of the power spectrum
  $\langle |h_\VECq^2|\rangle$ for the bilayer system with
  $w_\romc/\sigma=1.4$ and $k_\romB
  T/\epsilon=1.0$.}\label{fig:modeanalysis}
\end{figure}

After setting up the bilayer, we first waited until tension, box
length, and energy had equilibrated (which took typically $10^5
\,\tau$ for fluid systems).  Then on the order of 100 configurations
separated by $500\,\tau$ were used to measure the mode spectrum.  The
bilayer mid-plane was identified by tracking the tail-beads and
interpolating their vertical position onto a $16\times 16$ grid.
Possible stray lipids had to be excluded from this procedure.  A Fast
Fourier Transform then yields the power spectrum $\langle
|h_\VECq^2|\rangle$, but this requires one additional correction due
to amplitude under-sampling on a grid which we briefly discuss in
Appendix \ref{app:spectral_damping}.  Fig.~\ref{fig:modeanalysis}
provides a typical example of such a mode spectrum from which we can
clearly see the asymptotic $q^{-4}$ scaling, but also the deviations
at large $q$.  In this case length scales exceeding $L\approx
20\,\sigma$ (\ie, about four times the bilayer thickness) are required
to reach the asymptotic regime.  Hence, a simulation of smaller
systems (1000 lipids, $L\approx 25\sigma$) would not suffice to obtain
a fluctuation spectrum which could in any meaningful way be fitted to
Eqn.~(\ref{eq:hq2}).

\subsubsection{Diffusion constant}

Calculating the in-plane diffusion constant $D_\romb$ for lipids in a
solvent-free bilayer system is not as trivial a task as it might at
first seem.  For a strictly two-dimensional diffusive system we would
of course have
\begin{equation}
  D_\romb = \frac{1}{4\Delta t}\Big\langle \big[\VECr_{||,i}(t+\Delta t) - \VECr_{||,i}(t)\big]^2 \Big\rangle_i \ ,
  \label{eqn:diff}
\end{equation}
where $\VECr_{||,i}(t)$ is the projection \footnote{Just as in a typical experiment we measure the effective diffusion in the projected plane, which is not the same as the bare diffusion in the bilayer surface because of membrane undulations.  For a discussion see Ref.~\cite{reister_seifert:2005}} of the
position vector of the $i^{\text{th}}$ lipid at time $t$ into the
bilayer plane, $\Delta t$ is the time difference over which diffusion
is probed, and the angular brackets indicate an average over all $N$
lipids as well as configurations separated by a time difference of
$\Delta t$.  The difficulty, though, is that the system is really
three-dimensional and that a small but nonvanishing fraction $\alpha
\ll 1$ of all lipids (typically $\alpha\approx 1-2\,$\%) resides in
the gas phase surrounding the bilayer.  Even though they are the
minority phase, they of course diffuse much faster and might thus
contribute significantly to the average mean squared horizontal
distance traveled.  Unfortunately it is not straightforward to
eliminate such stray lipids from the average $\langle\cdots\rangle_i$
over all lipids in Eqn.~(\ref{eqn:diff}), since this would require us
to check on the positions of all lipids during all times
\emph{between} $t$ and $t+\Delta t$ --- information that is not
necessarily available.

\begin{figure}
\includegraphics[scale=0.78]{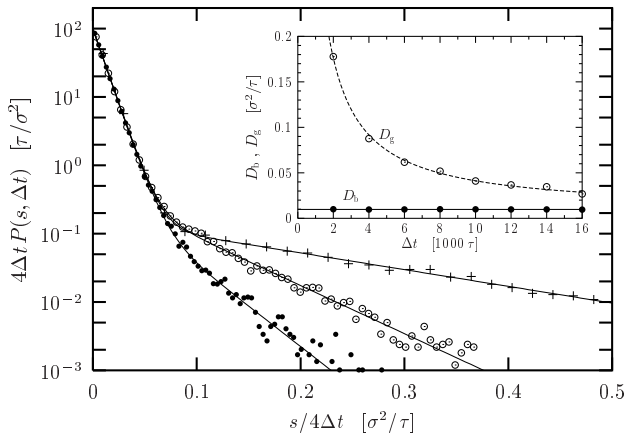}
\caption{Scaled distribution of squared displacements $s$ of lipid
molecules for the system with $w_\romc/\sigma=1.6$, $k_\romB
T/\epsilon=1.1$ and $\Delta t = 2000\,\tau$ (crosses), $\Delta t =
6000\,\tau$ (open circles), and $\Delta t = 12000\,\tau$ (filled
circles).  The lines are fits to Eqn.~(\ref{eqn:prob}).  These always
yield $\alpha\approx 0.02$ and values of the two diffusion constants
as given in the inset.  While $D_\romb$ is constant, the data are
compatible with $D_\romg$ approaching $D_\romb$ with an $1/\Delta t$
asymptotics (inset, dashed
line).}\label{fig:displacement_distribution}
\end{figure}

Fortunately there is a way to check Eqn.~(\ref{eqn:diff}) which does
not require us to follow lipids, namely, by looking at the entire
distribution function $P(s)$ of lipid displacements $s=(\Delta
\VECr_{||})^2$.  If we think of lateral lipid diffusion as being
due to two independent simple diffusion processes -- one with bilayer
diffusion constant $D_\romb$ and one with an effective lateral
diffusion constant $D_\romg\gg D_\romb$ through the gas phase -- a
simple calculation would suggest
\begin{equation}
P(s,\Delta t)
=
(1-\alpha)\frac{\rome^{-s/4D_\romb \Delta t}}{4D_\romb \Delta t} 
 + \alpha \frac{\rome^{-s/4D_\romg \Delta t}}{4D_\romg \Delta t} \ .
\label{eqn:prob}
\end{equation}
Indeed, a histogram of squared traveled distances shows exactly this
double exponential decay, with the short distance behavior dictated by
the real bilayer diffusion constant $D_\romb$ (see
Fig.~\ref{fig:displacement_distribution}).  We tested in different
cases (including ones with a rather large fraction of stray lipids)
that $D_\romb$ is independent of $\Delta t$ and in fact given by the
diffusion constant obtained from the much simpler analysis via
Eqn.~(\ref{eqn:diff}).  In contrast, $D_\romg$ depends essentially
inversely on $\Delta t$, which rather than suggesting that diffusion
through the gas phase somehow slows down reminds us that free
diffusion through three-space is ultimately cut short via readsorption
in the bilayer or its periodic image.

Having thus established that fast moving stray lipids have no
significant influence on the bilayer diffusion constant as measured
via Eqn.~(\ref{eqn:diff}), we subsequently used this simpler analysis
to obtain $D_\romb$.  In order to make best possible use of our data
we calculated $D_\romb$ by taking the average over all possible values
of the starting time $t$ for each interval length $\Delta t$ (which
varied from $2000\,\tau$ up to the entire length of an equilibrium
simulation run ($\sim 100000 \tau$)) and then took the weighted
average over all possible values of $\Delta t$.

\subsubsection{Flip-Flop rate}

A lipid molecule will not stay forever in the particular monolayer in
which it presently resides; rather, there is a particular probability
per unit time, the flip-flop-rate $r$, that it changes the monolayer.
Let $N_+(t)$ and $N_-(t)$ be the number of lipids in the upper or
lower monolayer at time $t$, which were present in the \emph{upper}
layer at time $t=0$.  These numbers satisfy the Master equations
\begin{equation}
  N_\pm(t+\romd t) = N_\pm(t)\,(1-r\,\romd t) + r\,\romd t \, N_\mp(t) \ ,
\label{eq:master_flipflop}
\end{equation}
The total number of such lipids is conserved, $N_+(t)+N_-(t)=N/2$, so
we obtain $\dot{N}_\pm(t)=-r[2N_\pm(t)-N/2]$.  The solution satisfying
the initial conditions $N_+(0)=N/2$ and $N_-(0)=0$ is $N_\pm(t) =
\frac{1}{4}N(1\pm\rome^{-2rt})$.  The same considerations hold for
lipids which at $t=0$ were in the lower monolayer.  Hence, the total
fraction $f(t)$ of lipids which at time $t$ reside in the same
monolayer as they did at time $t=0$ is given by
\begin{equation}
  f(t) = \frac{1}{2}\big(1+\rome^{-2rt}\big) \ .
\label{eq:fraction_flipflop}
\end{equation}
This fraction is easily measurable in simulations.  A fit to
Eqn.~(\ref{eq:fraction_flipflop}) then yields the flip-flop-rate $r$.
Notice that the probability density for flipping at time $t$ is given
by $r\,\rome^{-rt}$, and thus the average time between flip-flops is
$\langle t \rangle = 1/r$.


\subsection{Constant temperature cuts}\label{sec:const_T}

\begin{figure*}
\includegraphics[scale=0.78]{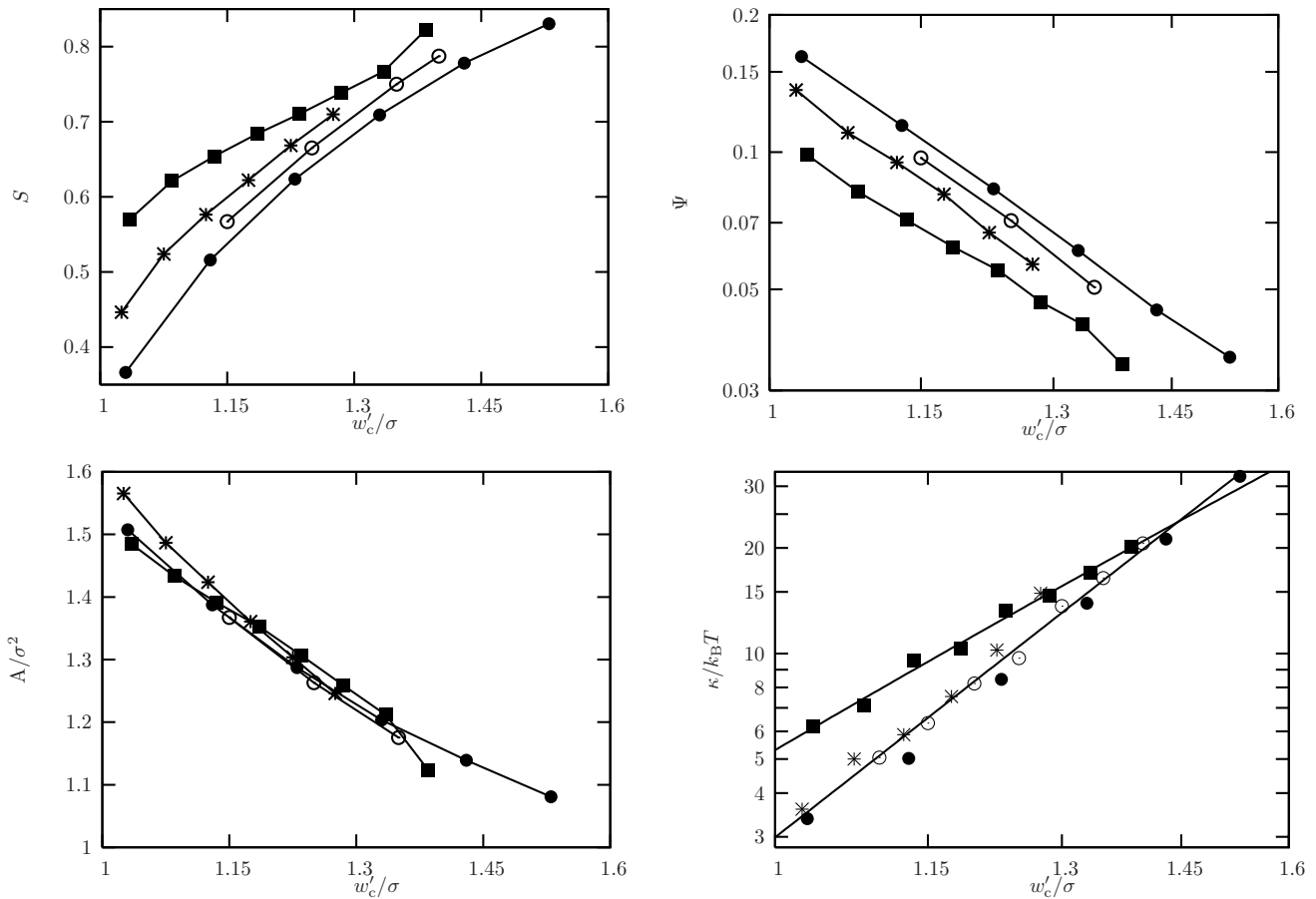}
\caption{Basic static properties of the fluid bilayer phase as a
function of $w_{\romc}^{\prime}$ and $k_\romB T$, where
$w_{\romc}^{\prime}$ indicates a rescaled attractive potential
($w_{\romc}^{\prime} = 1+ w_{\romc} - w_{\romc}^{\romlg} $). $w_{\romc}^{\romlg}$
is the value of $w_\romc$ on the liquid-unstable (gas) transition
line. Each plot shows four isotherms; $k_\romB T = 0.6 \epsilon$
(filled squares), $k_\romB T = 0.8 \epsilon$ (asterisks), $k_\romB T =
1.0 \epsilon$ (open circles), $k_\romB T = 1.1 \epsilon$ (filled
circles).  The values of $w_{\romc}^{\romlg}$ for each of these
isotherms were $0.185$, $-0.025$, $-0.2$ and $-0.27$ respectively. In
all cases statistical errors were smaller than the size of plotted
points.}\label{fig:lparams}

\end{figure*}

Fig.~\ref{fig:lparams} summarizes the values of orientational order
parameter $S$ (Eqn.~\ref{eq:s}), overlap order parameter $\Psi$
(Eqn.~\ref{eq:psi}), area per lipid $a$, and bending modulus $\kappa$ for
constant temperature scans along the isotherms $k_\romB T/\epsilon =
0.6$, $0.8$, $1.0$, and $1.1$.  In order to more directly compare
between results at different temperature we applied the rescaling
$w_{\romc}^{\prime} = 1 + w_{\romc} - w_\romc^{\romlg}$ where
$w_\romc^{\romlg}$ represents the value of $w_\romc$ at the liquid-gas
transition line for each respective temperature (we chose $1$ as a
reference value rather than $0$ to permit the construction of
log-scale plots) .  This allows us to scale out the most obvious
effects of temperature and brings the results for all of the isotherms
much closer to a common trend.

We first note that the trends for all observables are monotonic,
showing a consistent change from bilayers close to the liquid-gas
boundary to those close to the liquid-gel boundary.  Starting with the
structural parameters, $S$ and $\Psi$, we see that both indicate an
increase in order with larger potential range $w_\romc$.  This is not
immediately obvious, because broadening of the attractive potentials
could also give the lipids more lateral freedom.  Yet, the increase in
$S$ indicates that lipids fluctuate less around their average vertical
position, and this goes hand in hand with a concomitant decrease in
the overlap $\Psi$ between the vertical density profiles for
individual beads.  In fact, it turns out that the average bead
positions are roughly constant, hence the main mechanism by which
$\Psi$ can increase is via the broadening of $\rho(z)$ peaks for
individual beads (see Fig.~\ref{fig:zprof}).  Close to the liquid gas
boundary the bilayer order is rather low.  Indeed, visual inspection
of bilayers with $\Psi \gtrsim 0.1$ confirms that these are indeed
very ``fuzzy''.  Although a clear bilayer structure can still be seen
in such cases, we would caution against using these in the attempt to
model real lipid systems.

The increase in order becomes more understandable when looking at the
average area per lipid.  Extending the range of the cohesive potential
leads to an overall lateral contraction of the bilayer, thus
explaining the reduced fluctuations and thus the behavior of $S$ and
$\Psi$.  It is  remarkable that  the area per lipid of all four
isotherms  agrees after rescaling.  This indicates
that the lipid density depends purely on the distance from the
liquid-gas phase boundary rather than the absolute value of
temperature or $w_\romc$ individually.

The area $a$ per lipid can also be used to map the coarse-grained
length scale $\sigma$ to experimental lengths. For real phospholipid
membranes values around $0.75\,\text{nm}^2$ for the area per lipid are
typical \cite{shinoda_etal:1997,husslein_etal:1998}, while our
simulations give values in the range $1.1-1.5\,\sigma^2$.  Assuming
that one coarse-grained lipid is equivalent to one real lipid this
gives a mapping of roughly $\sigma \simeq 0.7-0.8\,\text{nm}$.  An
alternative mapping can be obtained by comparing a typical bilayer
thickness of roughly $5\,\text{nm}$ with the measured width of the
overall lipid density of approximately $5\,\sigma$ (see
Sec.~\ref{sec:cross_bilayer} and Fig.~\ref{fig:zprof}), which gives
$\sigma \simeq 1\,\text{nm}$.  Such good agreement between these two
mappings indicates that our very simple 3 bead lipids are actually
remarkably close to the aspect ratio of real lipids.

The observables presented so far reflect mostly local bilayer
properties.  In contrast, the bending stiffness is an observable
which, even though it ultimately derives from local bilayer
properties, yields physics which is accessible by large scale
continuum calculations.  Much of the theoretical modeling of fluid
membranes hinges on the remarkable fact that on sufficiently
large length scales they can be described by idealized surfaces with a
very simple energy density, for which the bending stiffness $\kappa$
is in almost all cases the only relevant modulus
\cite{canham:1970,helfrich:1973,seifert:1997}.  Reproducing
experimentally meaningful and easily tuneable values for this modulus
is therefore one of the key requirements for any coarse-grained
membrane modeling that aims at bridging the gap between local and
global scales.

Using the procedure described in Sec.~\ref{sec:bending_modulus} we
calculated $\kappa$ for each of the four isotherms $k_\romB T/\epsilon
= 0.6$, $0.8$, $1.0$, $1.1$ and for all values of $w_\romc$ 
in the fluid phase.  The two most important conclusions from these
data are the following: First, the range of accessible values for the
bending stiffness coincides exactly with the experimentally
interesting range for usual phospholipids \cite{seifert_lipowsky:1995}.  Second, the
value of $\kappa$ can be easily tuned via one parameter, the potential
range $w_\romc$.  Beyond that, it is quite remarkable that plotting
$\kappa/k_\romB T$ against the rescaled potential width $w_\romc'$
collapses all data points onto one master curve -- with the notable
exception of $k_\romB T/\epsilon = 0.6$, which also stands out for the
diffusion constant (see below).  As a guide to the eye, a single line
is shown for the three warmest isotherms and a separate line for the
coldest one.  Just as for the area per lipid it seems thus that the
bending modulus depends only on the distance from the liquid-gas
boundary.  This is plausible, since the bending modulus is
inversely proportional to the lateral compressibility, which again
depends on the lipid density.

\begin{figure}
\includegraphics[scale=0.8]{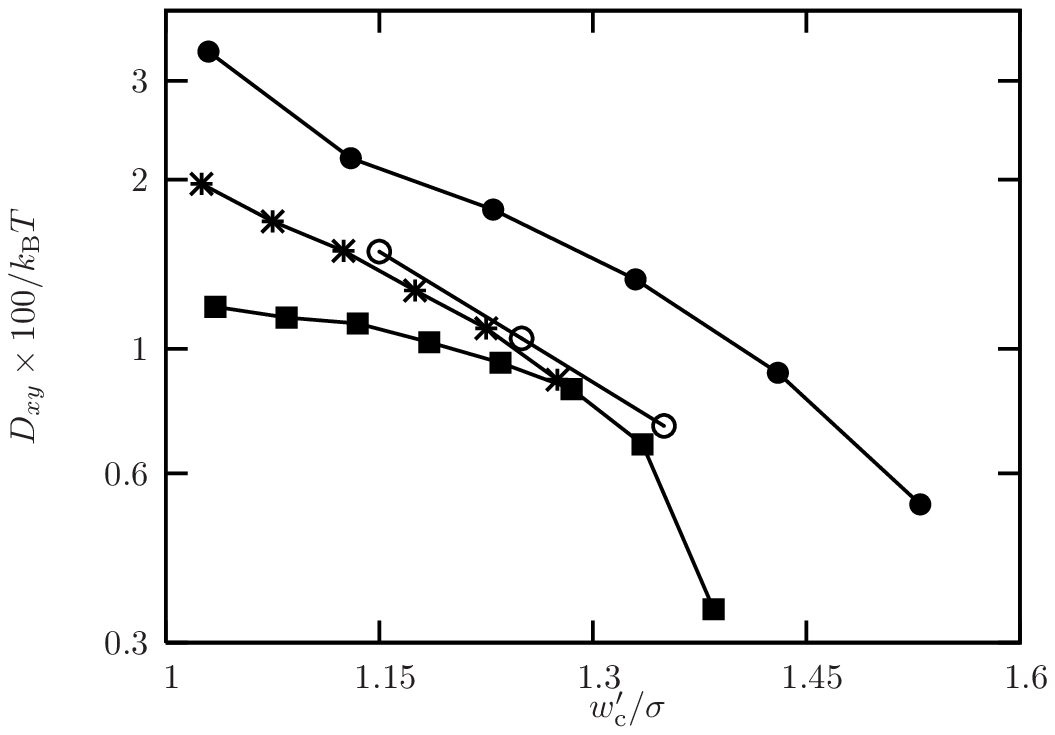}
\caption{Diffusion constant as a function of rescaled potential width
  $w_{\romc}^{\prime}=1+w_\romc-w_\romc^\romlg$.  Symbols and shifts
  $w_\romc^\romlg$ are the same as in
  Fig.~\ref{fig:lparams}.}\label{fig:diffcombo}
\end{figure}

\begin{figure}
\includegraphics[scale=0.8]{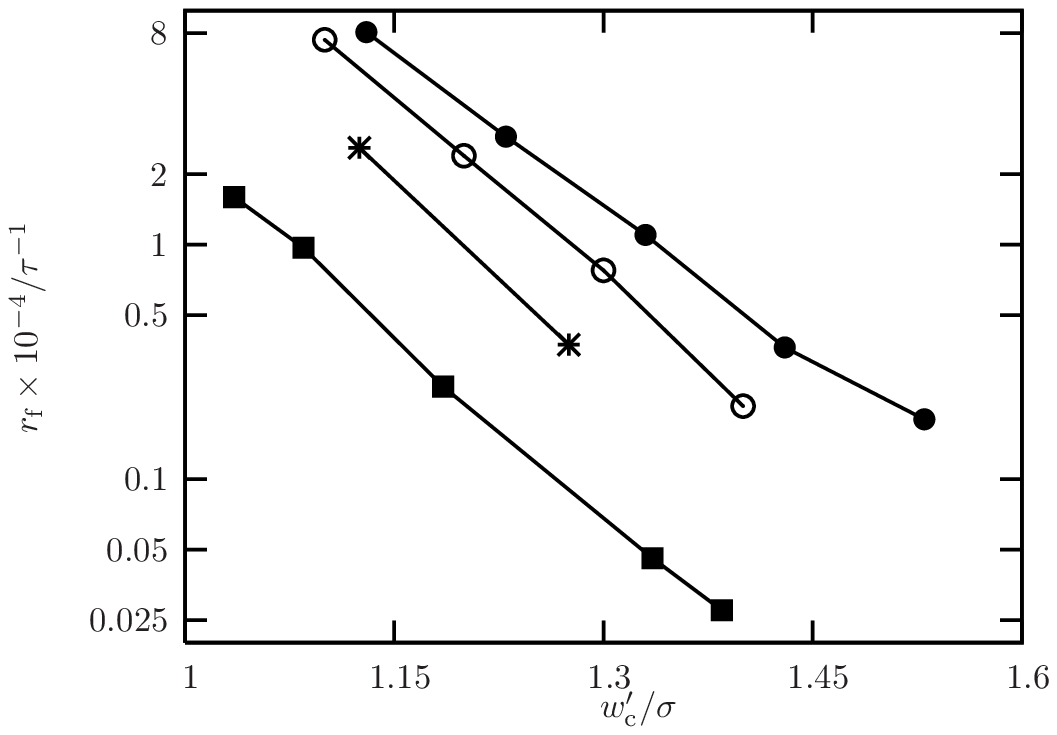}
\caption{Flip-flop-rate $r$ as a function of rescaled potential width
  $w_{\romc}^{\prime}=1+w_\romc-w_\romc^\romlg$.  Symbols and shifts
  $w_\romc^\romlg$ are the same as in
  Fig.~\ref{fig:lparams}.}\label{fig:rcombo}
\end{figure}

Both dynamical properties, the bilayer diffusion constant $D_\romb$
(Fig.~\ref{fig:diffcombo}) and the flip-flop-rate $r_\romf$
(Fig.~\ref{fig:rcombo}) show a clear exponential decay with increasing
$w_\romc^{\prime}$.  Since we know that for free diffusion $D \propto
T$, a simple rescaling by temperature brings the diffusion constants
for all isotherms into rough agreement, with the notable exception of
$k_\romB T/\epsilon = 0.6$.  We have seen that this same isotherm also
gives an anomalous trend in the bilayer stiffness, and one can
speculate that this may be due to the unusual shape of the phase
diagram (see Fig.~\ref{fig:cospd}) between $w_\romc/\sigma = 0.8$ and
$w_\romc/\sigma = 1.1$.

A typical value for the diffusion constant of lipids in real
phospholipid membranes is about $1\,\mu\text{m}^2/s$
\cite{fahey_webb:1978}.  Taking an average value of about
$0.01\,\sigma^2/\tau$ from our data, and using the length mapping
$\sigma\approx 1\,\text{nm}$ (see above) we obtain a time scale
mapping of $\tau\approx 10\,\text{ns}$.  Although we do not place
great quantitative weight on such an approximate calculation, it does
serve to illustrate that the timescales accessible by us are extremely
long, being of the order of milliseconds.  This is long enough, for
example, to allow vesicles to self assemble or fuse, and for
macroscopic phase separation to take place.

In the case of flip-flop rates we found a strong dependence on
temperature which is in accordance with the fact that this is a
thermally activated process.  The most important point to note however
is that the range of actual values for $r_\romf$ obtained by us are many
orders of magnitude faster than those typically found for artificial
phospholipid bilayers in experiments
\cite{liu_conboy:2005,kol_etal:2001}.  Such a large discrepancy is not
as alarming as it may at first seem since the rate $r_\romf$ is
exponentially dependent on the activation energy for flip $E_\romf$
(Arrhenius Law). We can therefore account for a very large difference
in $r_\romf$ by a relatively small discrepancy in $E_\romf$.  Nevertheless,
it is clear that the lipids in our model undergo flip-flop too easily
compared with real lipids.  If one were specifically interested in
this aspect of the dynamics, this would clearly represent a problem,
however in many other cases it provides an advantage because the
system will approach equilibrium more rapidly.  If an
accurate flip-flop rate is important we anticipate that our model
lipids could be made to flip much less readily simply by increasing
the chain length slightly (eg 4 bead lipids) and imposing a much
stronger head-tail repulsion.


\subsection{Constant $w_\romc$ profiles: Gel-Fluid Transition}\label{subsec:gelfluid}

\begin{figure}
\includegraphics[scale=0.75]{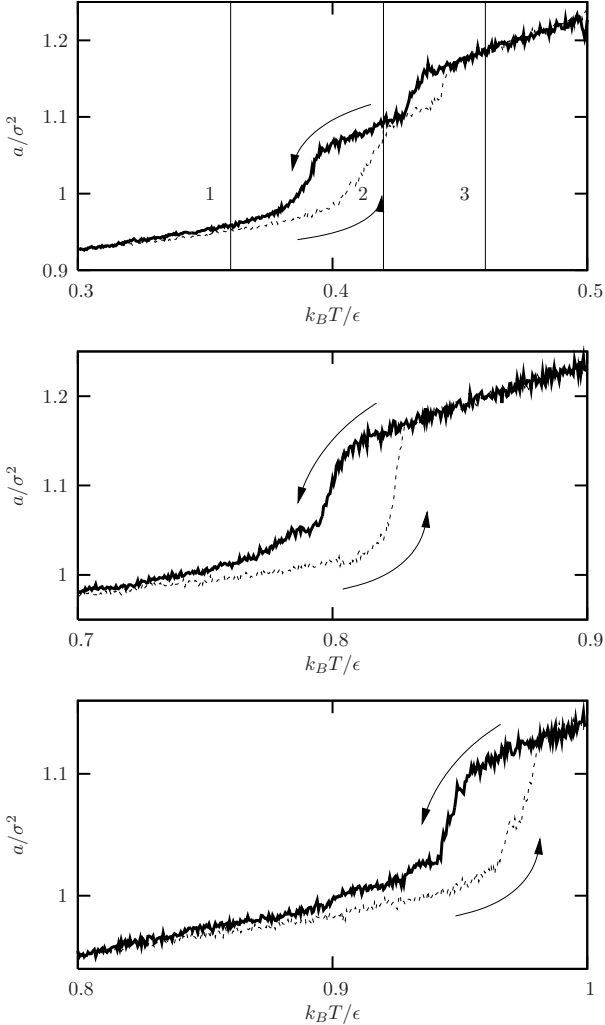}
\caption{Variation of the area per lipid $a$ across the gel-fluid
  phase boundary. Each figure shows a cooling-heating hysteresis for a
  particular value of $w_\romc$.  From top to bottom the values of
  $w_\romc/\sigma$ used were $1.0$, $1.4$, and $1.6$.  Arrows indicate
  the direction of temperature change.  The rate of temperature change
  was $2.5 \times 10^{-6}\,\epsilon/k_\romB\tau$ for the top plot and $5
  \times 10^{-6}\,\epsilon/k_\romB\tau$ for the bottom two plots.  The
  three vertical lines in the uppermost plot indicate the temperatures
  where the order parameter of Fig.~\ref{fig:lh} has been
  measured.}\label{fig:glboxl}
\end{figure}

At constant values of $w_\romc$ one can vary the temperature to
observe both liquid-gel and liquid-gas transitions.  Since we studied
lipid phase behavior at vanishing lateral tension, the liquid-gas
boundary is necessarily sharp in our case.  However, it is worth
mentioning that under constant volume conditions alternative low
density phases such as spherical or cylindrical micelles can be
observed.  Using the present model we have in fact observed such
micellar phases, and together with one more control parameter --
overall lipid density -- their phase behavior could be studied as
well.  This, however, is not the purpose of our present work and will
be presented elsewhere.  Here we concentrate on the transition
between liquid and gel phases at vanishing lateral tension.  There are
 several different ordered phases which collectively can be
referred to as gels
\cite{kranenburg_smit:2005,lipowsky_sackmann:1995}, but we shall not
attempt to identify all of these.  Indeed, our rather simple three
bead model was neither designed to reproduce such subtleties of lipid
ordering nor would we actually expect to observe the full zoo of
ordered bilayer phases.

In Fig. \ref{fig:glboxl} we show the variation of lipid area $a$ with
temperature for three values of $w_\romc$.  In all cases we chose to
employ continuous temperature scans in order to allow the barostat to
smoothly follow the  substantial changes in box size involved.
The important question then arises: ``is our rate of cooling slow
enough?''.  We checked this by comparing plots of $a$ \emph{vs}
$\tau$ for runs with four different cooling rates $10^{-4}$,
$10^{-5}$, $2.5\times 10^{-6}$, and $10^{-6}$ (in units of
$\epsilon/k_\romB\tau$), from which one can see a clear deviation for
the fastest rate but qualitatively very similar results for all slower
ones.  The only difference is that slightly sharper transition
boundaries can be seen as the rate is lowered, even among the slowest
three rates.  As a compromise between speed and resolution we chose
the intermediate rate $5\times 10^{-6}\,\epsilon/k_\romB\tau$ for runs
with $w_\romc/\sigma = 1.4$ and $w_\romc/\sigma = 1.6$ which display
only a single transition.  For runs with $w_\romc/\sigma = 1.0$, where
two transitions must be resolved, we chose the slower rate, $2.5\times
10^{-6}\,\epsilon/k_\romB\tau$.

Before focusing on the results at each value of $w_\romc$ it is
interesting to note that all runs show quite a strong hysteresis
across the transition boundary as is typical of first order
transitions. However, there is also a long tail to the hysteresis which appears
during cooling.  This is most likely due to the fact that the kinetics
during gelling is strongly determined by the slow healing of defects.
Indeed, we often observed such defects, many of which could be seen to
dissappear during further slow cooling.

\begin{figure}
\includegraphics[scale=0.75]{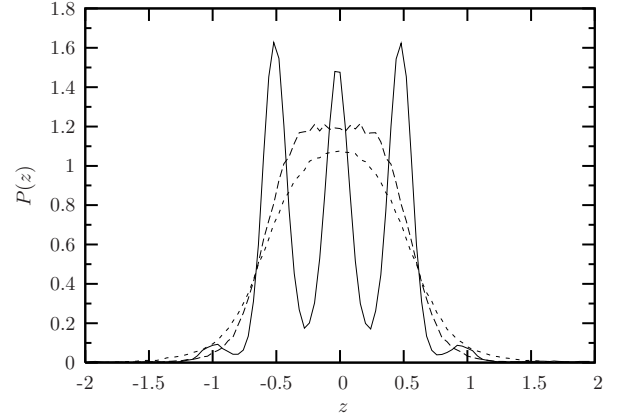}
\caption{Probability density $\rho(z)$ of the height difference $z$
between a lipid and its $6$ immediate neighbours.  Solid, dashed, and
dotted curves correspond to the temperatures indicated by lines 1, 2,
and 3 in Fig.~\ref{fig:glboxl}, respectively.} \label{fig:lh}
\end{figure}

Turning now to the results at $w_\romc = 1.0$ we find that there are
\emph{two} clear and sudden transitions during cooling as well as the reverse
for heating.  During cooling, both transitions involve a contraction
of the area $a$ and an increase in the orientational order $S$ (not
shown); however, if we look at the diffusion constant $D_\romb$ we
find that it decreases suddenly from about $2\times
10^{-3}\,\sigma^2/\tau$ to approximately $4\times
10^{-5}\,\sigma^2/\tau$ in the first (higher temperature) transition
but does not decrease further during the second (lower temperature)
transition.  In fact, the very small value of the diffusion constant
is numerically hard to determine accurately, but the overall drop by
about two orders of magnitude is probably robust and corresponds well
to what is known for typical phospholipid bilayers
\cite{fahey_webb:1978}.  Thus we have a first transition from fluid to
gel and a second transition at a lower temperature to a different
gel phase. In order to understand what is actually occurring in the
gel-gel transition we require a suitable order parameter, and it turns
out that local lipid packing is what one has to look at.  We
calculated a histogram of height differences $\Delta z$ between a
lipid head and that of its six nearest neighbors.  The results are
shown in Fig. \ref{fig:lh} for bilayers at temperatures $k_\romB
T/\epsilon = 0.38$, $0.42$, and $0.46$ corresponding to each of the
three phases.  Both the fluid phase and the high temperature gel phase
show a distribution peaked about $\Delta z=0 $; however, there is a
striking change as we go to the phase at $k_\romB T/\epsilon = 0.38$
which shows three peaks at $\Delta z = \pm 0.5\,\sigma$ and $0$.  This
shows that the lipids are now packed in an off-centered manner rather
like oranges in a crate.  Puzzlingly, however, an equal proportion of
lipids are found at $\Delta z = 0$ which would not occur if a perfect
packing was obtained.  This might be due to the presence of a still
large number of defects.

At larger values of $w_\romc$ there is just a single transition from
the fluid to the gel phase.  This gel phase is packed similarly to the
low temperature gel phase for $w_\romc = 1.0$ since it also exhibits a
three peak structure in the histogram of $\Delta z$.

Finally, the plots of area per lipid \emph{vs} temperature also
contain information besides the evident phase transitions.  Their
\emph{slope} yields the value of the lateral thermal area expansivity,
defined by
\begin{equation}
\alpha_T = \frac{1}{\langle a\rangle}\frac{\partial\langle a\rangle}{\partial T} \ .
\end{equation}
For instance, the system with $w_\romc/\sigma=1.6$ has a slope
$\partial\langle a\rangle/\partial T \approx
0.766\,\sigma^2/(\epsilon/k_\romB)$ in the fluid phase and an area per
lipid of $1.142\,\sigma^2$ at $k_\romB T/\epsilon=1$.  Identifying
this temperature with room temperature, we obtain a thermal
expansivity of $\alpha_T\approx 2.2\times 10^{-3}\,K^{-1}$, which
coincides remarkably well with the value measured by Kwok and Evans
for fluid lecithin bilayers ($2.4\times 10^{-3}\,K^{-1}$)
\cite{kwok_evans:1981}.


\section{Membrane Stretching and pore opening}\label{sec:stretch}
So far we have studied membranes under zero lateral tension.  If we
now apply extensional stress, the area per lipid will increase, up to
the point where structural stability of the bilayer is breached.
Beyond a critical stress a pore can be nucleated, which then grows
indefinitely, \ie, the bilayer ruptures.  This scenario has been
described theoretically by Litster \cite{litster:1975}.  Important
bilayer properties (\eg\ the line tension of an open edge) could be
extracted from observables such as the critical stress or the critical
pore size, but it is evidently experimentally very difficult to study
membranes at the brink of rupture. However, in a simulation it is very
easy to perform measurements in a different ensemble, namely one of
constant \emph{area} of the entire bilayer.  Beyond some critical
strain one would now expect a pore to open, but since this now
relieves much of the stress, pore growth stops and one obtains a
\emph{stable} pore of well-defined size.  Assuming a harmonic
extensibility of the bilayer itself, as well as a constant line
tension at a pore rim, Farago \cite{farago:2003} Tolpekina
\etal\ \cite{tolpekina_etal:2004} gave a simple theoretical model
which relates the resulting pore size as well as the stress-strain
relation to key bilayer properties such as the extensibility and the
line tension.  We summarize the key results in
Appendix~\ref{app:tolpekina}.

\begin{figure}
\includegraphics[scale=0.8]{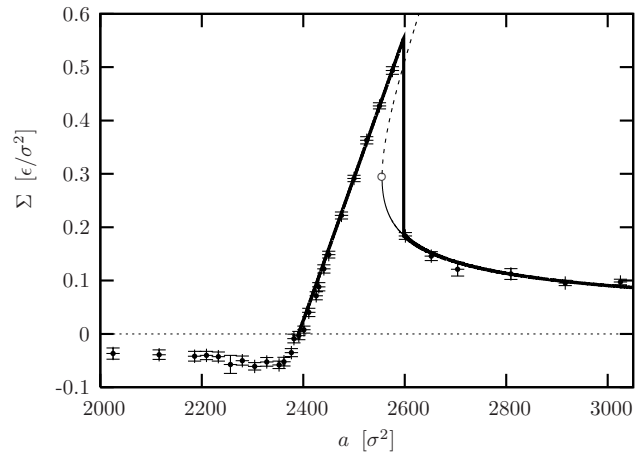}
\caption{Bilayer tension $\Sigma$ as a function of (projected)
area $A_\romtot$ for a flat membrane sheet with $w_\romc/\sigma = 1.6$
at $k_\romB T/\epsilon = 1.1$.  The bold solid line is a fit to the
model of Tolpekina \etal\ \cite{tolpekina_etal:2004} (see also
Eqn.~(\ref{eq:tolpekina_tension}) in Appendix~\ref{app:tolpekina});
the fine solid and dashed curves indicate metastable and unstable
branches, respectively.}\label{fig:tension}
\end{figure}

Using our simulation model with the parameters $w_\romc/\sigma = 1.6$
and $k_\romB T/\epsilon = 1.1$ we determined the equilibrium lateral
tension $\Sigma$ as a function of total box area ($A = L_x \times
L_y$).  For each value of the area this was done by placing 4000
lipids in a bilayer configuration spanning the $xy$-plane of the
simulation box.  After allowing for an equilibration time of $3 \times 10^{4}$ we
then simulated for a further $3 \times 10^{4}$, during which time the lateral
tension $\Sigma = -L_z (p_{xx} + p_{yy})/2$ was measured.  The
resulting plot of $A$ \emph{vs} $\Sigma$ is shown in
Fig.~\ref{fig:tension}.  Three main regimes are clearly
distinguishable, separated by two values of the bilayer area.  First,
at a particular area $A_0$ the tension vanishes.  Boxes with a smaller
area yield a negative tension, \ie, a positive pressure.  The bilayer
is under compressional strain, which it evades very soon by
\emph{buckling}.  Conversely, boxes with an area larger than $A_0$
subject the bilayer to extensional strain and create a proportional
increase in tension.  At some particular value for the bilayer area
the energy stored in the extension, which grows quadratically with the
strain, must exceed that of a bilayer with a pore, since the pore size
cannot grow faster than the strain, and the line tension is
proportional to the square root of pore size.  Once the pore opens it
releases much of the stress and further expansion of the area
remarkably leads to further {\it decrease} in the stress.  This
``wrong'' sign for the extensibility explains why a stable pore cannot
be achieved under constant tension conditions.

The model of Farago \cite{farago:2003} and Tolpekina \etal\ gives
perfect agreement with the measured data \footnote{There are 3 fitting parameters; zero tension area $A_0$, stretching modulus $M$,  and line tension $\gamma$ (see Appendix B).  These can be obtained purely from the initial stretching regime and the point of rupture.   Notably, the subsequent drop in tension and the shape of the curve after pore opening are then completely determined and not subject to further fitting. } (see lines in
Fig.~\ref{fig:tension}).  This, however, not only shows that their
theoretical assumptions were correct but also provides a reference for
comparison of our model with the well known model of Goetz and
Lipowsky \cite{goetz_lipowsky:1998,goetz_etal:1999} since Tolpekina
\etal\ fitted their results to this earlier model.  Contrary to ours,
the model of Goetz and Lipowsky includes explicit solvent.  That both
simulations can be described very well by the same theory implies that
their physical behavior under extensional stress is indeed very
similar.

Fitting the linear extensional regime of our simulational data to the
stretching model we obtain a modulus of $M\approx
6.4\,\epsilon/\sigma^2$.  Translating this to real values depends
again on the mapping.  Using $1.1\,\epsilon = k_\romB T =
4.1\,\text{pN}\,\text{nm}$ as well as a length mapping of
$\sigma\approx\,0.9\text{nm}$ gives $M\approx 30\,\text{mN}/\romm$.  This
is at the lower end of what one typically expects for phospholipid
bilayers \cite{kwok_evans:1981,evans_rawicz:1990,rawicz_etal:2000}.  The rupture tension is
$\Sigma_{\text{pore}}\approx 0.55\,\epsilon/\sigma^2$, which
translates to about $2.5\,\text{mN}/\text{m}$, while the line tension
is $\gamma\approx 1.2\,\epsilon/\sigma$, giving about $5\,\text{pN}$.
Both these values are very reasonable \cite{evans_etal:2003,loi_etal:2002}, but one has to be
careful, since they depend in a nontrivial way on system size
\cite{tolpekina_etal:2004}.


\section{Conclusions}

We have presented a model for the solvent free simulation of coarse
grained lipid bilayer membranes that is free of the major problems
encountered by earlier efforts towards this goal.  It robustly
self-assembles to a fluid phase, uses simple two body potentials, and
is highly tuneable.  We have emphasized that we regard its functioning
to rely on a general principle: the presence of sufficiently broad
tail attraction potentials.  This gives lipids a chance of lateral
mobility while maintaining fluidity.  Since rearrangements and their
concomitant local increase in pair distances do not immediately cost
most of the binding energy, the entropy gain upon fluidization is not
inhibited energetically.  In this context we remind the reader that it
is a well known fact from colloidal physics that if the range of
attraction is too short compared to the particle's hard core
radius, no more fluid phase exists \cite{gast_etal:1983,hagen_frenkel:1994,louis:2001}.  This principle should
therefore also help if extensions to our present model, such as more
beads per lipid or a restriction of the long-ranged attraction to
specific beads only are to be employed.

We have measured many physical characteristics of our membrane model
and have illustrated that they are easily tuneable in a controlled
way.  Often their values fall well inside or close to the
experimentally interesting range without any explicit careful tuning.  This indicates that the model is clearly flexible
enough and can serve as a good starting point for quantitative
matching to specific systems in the spirit of systematic coarse
graining, which is an approach that has only recently been employed for lipid membrane simulations \cite{shelley_etal:2001,izvekov_voth:2005} but is well established in other fields of soft matter \cite{mplathe02,praprotnik_etal:2005} .  Since we have demonstrated that we can readily
achieve mesoscopic length scales beyond $100\,\text{nm}$ and time
scales of milliseconds, this opens up a wide range of interesting
mesoscale problems that can with some additional parameter matching be
simulated quantitatively.  Efforts in this direction are currently
under way.

\acknowledgments

We thank Oded Farago, Friederike Schmid, Hiroshi No\-gu\-chi, Gregoria
Illya, Kurt Kremer, and Bernward Mann for valuable discussions.


\appendix


\section{Spectral damping}\label{app:spectral_damping}

In Sec.~\ref{sec:bending_modulus} we determined the bending modulus
from the membrane fluctuation spectrum, which in turn was extracted
from the simulation by interpolating the lipid positions onto a
$16\times 16$ grid.  However, any method which interpolates continuous
variables onto grid points is prone to discretization artifacts, and
it is crucial to identify them in advance.

We illustrate the difficulty in the one-dimensional case first.  Let
$h_q(x)$ be a single mode on the linear interval $[0;L]$, given by
\begin{equation}
  h_q(x) = \rome^{\romi (q x + \varphi_q)}
  \quad,\quad
  q = \frac{2\pi n}{L}
  \quad,\quad
  n \in \ZZ \ ,
\end{equation}
where $\varphi_q$ is a $q$-dependent phase.  On the grid this
mode has the values $H_q(k) = h_q(x_k)$ with $x_k = kL/N$ and $k \in
\{0,1,\ldots,N-1\}$.  The inverse Finite Fourier Transform $\hat{H}_q(u)$
of these sample points is given by
\begin{eqnarray}
  \hat{H}_q(u) & = & \frac{1}{N} \sum_{k=-\frac{N}{2}}^{\frac{N}{2}-1}
  H_q(k) \, \rome^{-2\pi\romi uk/N} \nonumber \\ & = & \frac{1}{N}
  \sum_{k=-\frac{N}{2}}^{\frac{N}{2}-1} \rome^{2\pi\romi
  (n-u)k/N}\rome^{\romi\varphi_k} \nonumber \\ & = &
  \rome^{\romi\varphi_k}\,\delta_{n,u} \ .
\end{eqnarray}
This is expected: If $n=u$ we get back the phase of our mode, but if
$n\ne u$ we get $0$, reflecting the fact that the wave numbers $u$ and
$n$ do not match.

The situation changes once we go beyond plain sampling.  In the
bilayer situation under study our mode $h_q(x)$ is being represented
by many off-lattice points $x\in[0;L]$.  Grid interpolation was
achieved by assigning to any grid point the average of the $h_q(x)$
for the $x$-values closest to that grid point.  If there are
many such points, this basically means that we do not sample the mode
at the grid point, but rather sample its \emph{average} around that
grid point.  Let us thus define the average-sampled interpolation
function $\bar{H}_q(k)$ via
\begin{eqnarray}
  \bar{H}_q(k)
  & = &
  \frac{1}{L/N}\int_{x_k-\frac{L}{2N}}^{x_k+\frac{L}{2N}} \romd x \; \rome^{\romi(qx + \varphi_q)}
  \nonumber \\
  & = &
  \frac{N}{\romi q L}\rome^{\romi(qkL/N+\varphi_q)}\Big(\rome^{\romi qL/2N}-\rome^{-\romi qL/2N}\Big)
  \nonumber \\
  & = &
  \sinc\Big(\frac{n}{N}\Big)\, H_q(k) \ ,
\end{eqnarray}
where we introduced $\sinc(x) = \sin(\pi x)/\pi x$.  Evidently
$\bar{H}_q(k)$ differs from the ordinary sampled interpolation
function $H_q(k)$ by an additional wave-vector dependent but position
independent prefactor.  This factor goes to zero at $n=N$, which
implies that such modes are entirely averaged away, but already at the
boundary of the Brillouin zone, at $n=\frac{N}{2}$, it has the
disconcertingly small value $\sinc\frac{1}{2} = \frac{2}{\pi} \approx
0.64$.  Once the mode has arrived on the mesh, it's amplitude is no
longer what it used to be off-lattice.

The inverse Finite Fourier Transform of $\bar{H}_q(k)$ is also
multiplied by this prefactor:
\begin{eqnarray}
  \hat{\bar{H}}_q(u)
  & = &
  \frac{1}{N} \sum_{k=-\frac{N}{2}}^{\frac{N}{2}-1} \sinc\Big(\frac{n}{N}\Big)\,H_q(k) \, \rome^{-2\pi\romi uk/N}
  \nonumber \\
  & = &
  \sinc\Big(\frac{n}{N}\Big)\,\hat{H}_q(u) \ ,
\end{eqnarray}
and thus the power spectrum is damped according to
\begin{equation}
  \big|\hat{\bar{H}}_q(u)\big|^2
  =
  \sinc^2\Big(\frac{n}{N}\Big)\,\big|\hat{H}_q(u)\big|^2
  =
  \sinc^2\Big(\frac{n}{N}\Big)\,\delta_{n,u} \ .
\end{equation}
At the boundary of the Brillouin zone this factor is $(2/\pi)^2\approx
0.41$ and thus not at all negligible.  Notice that this does not
depend on the mode amplitude.  In other words, even if the mode is
only slightly excited and the membrane thus looks benignly flat, the
estimate for the power spectrum derived from the gridded function is
systematically too low.

It is easy to see that in two dimensions this result generalizes to
\begin{equation}
  \big|\hat{\bar{H}}_\VECq(\VECu)\big|^2
  =
  \sinc^2\Big(\frac{n_x}{N}\Big)\,\sinc^2\Big(\frac{n_y}{N}\Big)\,\delta_{\VECn,\VECu} \ .
\end{equation}

The fact that this damping is $\VECq$-dependent renders it potentially
harmful.  Fig.~\ref{fig:modeanalysis} shows that a typical
zero-tension fluctuation spectrum at low values decays as $q^{-4}$ but
starts to bend upwards once the wave vector approaches microscopic
scales.  The spectral damping discussed above will act to
\emph{suppress} this upturn and delay it to even larger $q$-values.
The $q^{-4}$ regime might therefore appear to extend further than it
actually does, and one might thus be tempted to fit an expression
valid only in the bending regime to data which owe their exponent to a
combination of, say, protrusion modes and spectral damping artifacts.

In our mode-analysis we avoided such artifacts by simply
\emph{dividing out} the damping factor.


\section{Membrane pore in a constant-area-ensemble}\label{app:tolpekina}

For the total energy of an elastic bilayer spanned inside a frame of
area $A$ which displays a harmonic extensibility with modulus
$M$ and which has a line tension $\gamma$, \cite{farago:2003,tolpekina_etal:2004} write
\begin{equation}
  E = \frac{1}{2}M\frac{(A-A_0-\pi R^2)^2}{A_0} + 2\pi \gamma R \ ,
\end{equation}
where $A_0$ is the tensionless area of that bilayer and $R$ the radius
of a circular pore.  If $R$ is indeed nonzero, its value has to be
chosen such as to minimize $E$.

It is useful to rescale variables.  Let us define a characteristic length
scale $\lambda$ and with its help introduce a dimensionless pore
radius and area extension:
\begin{equation}
\lambda^3 = \frac{\gamma A_0}{\pi M}
\quad,\quad
\tR = \frac{R}{\lambda}
\quad,\text{ and}\quad
B = \frac{A-A_0}{\pi\lambda^2}
\label{eq:lambda}
\end{equation}
The equation for the optimal pore radius, $\partial E/\partial R=0$,
then reduces to
\begin{equation}
\tR^3-B\tR+1 = 0 \ .
\label{eq:pore_equi_2}
\end{equation}
Hence, the pore opening scenario will exclusively depend on only one
characteristic dimensionless variable, $B$; furthermore, all length
scales will be proportional to $\lambda$ with a prefactor that depends
on $B$ alone.

It is easy to see that $R=0$ is always a local minimum for $E$ (even
though the derivative does not vanish there).  For $B > B_\romc =
3/2^{2/3} \approx 1.89$ two more stationary radii appear as solutions
of Eqn.~(\ref{eq:pore_equi_2}):
\begin{equation}
\tR_\pm(B) = 2\sqrt{B/3}\,\cos\frac{\pi\pm\arctan\sqrt{4(B/3)^3-1}}{3} \ .
\end{equation}
The solution $\tR_-(B)$ corresponds to a local minimum and for large
$B$ asymptotically scales like $\sqrt{B}$; for $B > B_{\text{pore}} =
3/2^{1/3} \approx 2.38$ this minimum becomes the global one and a
discontinuous transition to a pore-state occurs, which displays a
system size dependent energy barrier of $E_{\text{barrier}}\approx
1.38\,\gamma\lambda \approx 0.94 \, \gamma^{4/3}M^{-1/3}A_0^{1/3}$.
In this pore state the tension is given by
\begin{equation}
\Sigma_-(B)
=
\frac{\gamma}{\lambda}\big[B-\tR_-^2(B)\big]
\stackrel{B\gg 1}{\simeq} \frac{1}{\sqrt{B}} + \mathcal{O}(B^{-2}) \ .
\label{eq:tolpekina_tension}
\end{equation}
At the transition point the tension drops exactly by a factor 3, and
the system size dependent rupture tension is given by
$\Sigma_{\text{pore}} = 3/2^{1/3}\,\gamma/\lambda \approx
2.38\,\gamma/\lambda \approx 3.49\,\gamma^{2/3}M^{1/3}A_0^{-1/3}$.



\end{document}